\documentclass[twocolumn, apj]{openjournal}

\def\gax{\mathrel{\raise.3ex\hbox{$>$}\mkern-14mu\lower0.6ex\hbox{$\sim$}}}
\def\lax{\mathrel{\raise.3ex\hbox{$<$}\mkern-14mu\lower0.6ex\hbox{$\sim$}}}
\def\gtorder{\mathrel{\raise.3ex\hbox{$>$}\mkern-14mu
             \lower0.6ex\hbox{$\sim$}}}
\def\ltorder{\mathrel{\raise.3ex\hbox{$<$}\mkern-14mu
             \lower0.6ex\hbox{$\sim$}}}

\usepackage{graphicx}
\usepackage{booktabs}
\usepackage{hyperref}
\usepackage{xspace}
\usepackage{orcidlink}

\hypersetup{
    colorlinks=true,
    linkcolor=blue,
    citecolor=blue,
    urlcolor=blue
}

\newcommand{\orcid}[2]{\href{https://orcid.org/#2}{\textcolor{blue}{#1}\,\includegraphics[height=8pt]{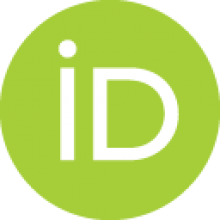}}}

\begin{document}

\title{Rotational Variables: \textit{Kepler} Versus ASAS-SN}

\author{
\orcid{Jack Stethem$^{1,\,\star}$}{0009-0006-1416-602X},
\orcid{Christopher S. Kochanek$^{1,2}$}{0000-0001-6017-2961},
\orcid{Anya Phillips$^{3}$}{0009-0005-1914-974X},
\orcid{Lyra Cao$^{4}$}{0000-0002-8849-9816},\\
and \orcid{Marc Pinsonneault$^{1}$}{0000-0002-7549-7766}
}

\affiliation{$^{1}$Department of Astronomy, The Ohio State University, 140 W 18th Ave, Columbus, OH 43210}
\affiliation{$^{2}$Center for Cosmology and AstroParticle Physics, 191 W Woodruff Ave, Columbus, OH 43210}
\affiliation{$^{3}$Center for Astrophysics \textbar\ Harvard \& Smithsonian, 60 Garden Street, Cambridge, MA 02138}
\affiliation{$^{4}$Department of Physics and Astronomy, Vanderbilt University, 6301 Stevenson Center Lane, Nashville, TN 37235}

\thanks{$^\star$Corresponding author: \href{mailto:stethem.5@buckeyemail.osu.edu}{stethem.5@buckeyemail.osu.edu}}

\begin{abstract}
Rotational variables are stars that vary in brightness due to star spots modulated by rotation. They are probes of stellar magnetism, binarity, and evolution. \cite{phillips2023} explored distinct populations of $\sim$ 50,000 high--amplitude rotational variables from the All-Sky Automated Survey for Supernovae (ASAS-SN), examining correlations between stellar rotation, binarity, and activity. Here, we carry out a similar analysis of $\sim$ 50,000 much lower amplitude \textit{Kepler} rotational variables. The \textit{Kepler} population is dominated by slowly rotating, single, main sequence stars, with a striking absence of the rapidly rotating main sequence group in the ASAS-SN sample. The binary fractions of the \textit{Kepler} rotators are significantly lower than for the ASAS-SN systems and they are significantly less spotted, as expected from their lower amplitudes. The scope of these statistical surveys will dramatically increase in the near future.
\end{abstract}

% \keywords{Classical Novae (251) --- Ultraviolet astronomy(1736) --- History of astronomy(1868) --- Interdisciplinary astronomy(804)}

\section{Introduction} \label{sec:intro}

% Rotating stars with convective envelopes generate magnetic fields which then produce
% star spots at the surface (see the review by \citealt{Strassmeier2009}).  Since the
% distribution of the spots is not uniform, the light curve of the star varies 
% with the period of rotation, leading to the class of rotational variable stars.
% Because the spot patterns evolve, the light curves also evolve and are not
% strictly periodic.  

Rotating stars with convective envelopes generate magnetic fields which then produce star spots at the surface (see the review by \citealt{Strassmeier2009}). Since the distribution of these spots is not uniform, the light curve of the star varies with the period of rotation, giving rise to the class of rotational variable stars. Because the spot patterns evolve over time, the light curves also evolve and are not strictly periodic.
Magnetically active main sequence stars lie at temperatures
below the \cite{Kraft1967} break where the convective zones of hotter stars becomes too
thin to support significant magnetic activity and winds.  
For these active stars, the magnetic activity then declines with age because
angular momentum loss through winds slows the rotation (\citealt{Skumanich1972}).   

This leads to a trend of slower rotation for greater ages that can be used
to estimate stellar ages through ``gyrochronology'' (e.g., \citealt{Epstein2014},
\citealt{Angus2019}, \citealt{Bouma2023}).  Stars in
sufficiently close binaries avoid spinning down by becoming tidally locked at the 
orbital period and can maintain their magnetic activity throughout their
main sequence life time (\citealt{wilson1966}).
For binaries with sufficiently short orbital periods, continued angular momentum loss
through stellar winds can lead to a merger and a rejuvenated, rapidly
rotating single star (\citealt{andronov2006}). 

Evolved stars should generally rotate slowly and be relatively inactive because
of both the earlier angular momentum losses and the consequences of expansion 
and angular momentum conservation.  Binary tidal interactions and mergers can
lead to faster rotation and more activity (see, e.g., \citealt{massarotti2008},
\citealt{carlberg2011}, \citealt{tayar2015}, \citealt{Daher2022}, \citealt{Patton2023}).  
Known populations of rapidly rotating, synchronized binaries are the 
RS~CVn stars (\citealt{hall1976}) and the sub-subgiants (\citealt{leiner2022}).

\cite{phillips2023} analyzed $\sim 50,000$ high-amplitude
($\gtorder 0.03$~mag) rotational variables in the All-Sky Automated Survey for Supernovae
(ASAS-SN, \citealt{Shappee2014}, \citealt{Kochanek2017}).  
These included previously known systems and systems identified by
searches for variable stars in ASAS-SN \citep{jayasinghe2018, jayasinghe2019I, jayasinghe2019II, 
jayasinghe2020, jayasinghe2021, christy2023}.   
\cite{christy2023} noted that the rotational variables lay in distinct clusters  
in the space of period and absolute magnitude, and
\cite{phillips2023} 
explored this in detail.  Using data from
APOGEE (Apache Point Observatory Galactic Evolution Experiment) DR17 (\citealt{apogeedr172022}) and \textit{Gaia} DR3 (\citealt{gaia2016}, \citealt{gaia2022}), 
they could determine whether the initial classes
consisted of single stars, binary stars or a mixture.  As in \cite{Daher2022}, the
binaries were largely identified by having a sufficiently high scatter in repeated
radial velocity (RV) measurements, although some orbital and rotational period comparisons
could be made for \textit{Gaia} DR3 stars with full orbital solutions \citep{Gosset2024}.  They also examined
the spot covering fractions using the method of \cite{cao2022}. 

\begin{figure*}
    \centering
    \includegraphics[width=.8\textwidth]{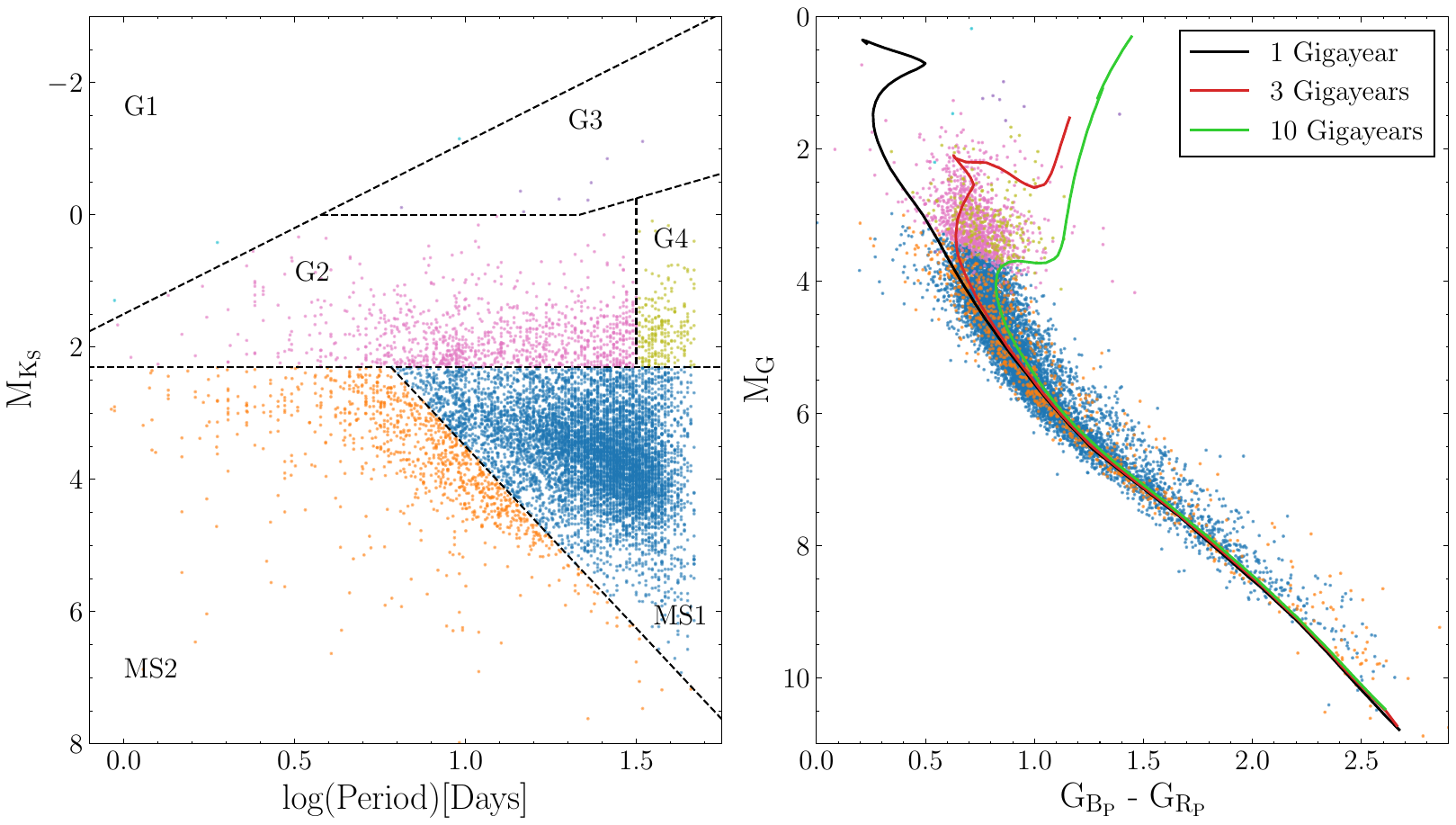}
    \caption{(Left panel) Distribution of a random 15 percent of the \textit{Kepler} stars in absolute $K_S$ magnitude and period along with the rotational variable class boundaries from \cite{phillips2023}. (Right panel) CMD of the same stars in absolute G magnitude and $B_P - R_P$ color, color-coded by class. The curves are 1, 3, and 10 Gyr Solar metallicity PARSEC \citep{parsec1, parsec2} isochrones.}
    \label{fig:cmd}
\end{figure*}

\begin{table*}[t]
\centering
\caption{Statistics of the Kepler sample \label{table1}}
\begin{tabular}{lrrrrrrr}
\toprule
 & Total & MS1 & MS2 & G1 & G2 & G3 & G4 \\
\midrule
Gaia & 55107 & 41510 & 5717 & 45 & 6529 & 30 & 1276 \\
Gaia w/ Orbital Solutions & 501 & 302 & 51 & 0 & 129 & 2 & 17 \\
APOGEE & 2496 & 1519 & 292 & 4 & 365 & 4 & 49 \\
APOGEE w/ Amplitudes & 906 & 496 & 199 & 1 & 107 & 1 & 4 \\
\bottomrule
\end{tabular}
\end{table*}

There were three
main sequence groups. MS1 consisted of relatively slowly rotating single stars
with periods of $P_{rot} \sim 10$-$30$~days. MS2s consisted of single stars that
are more rapidly rotating ($\ltorder 10 $days), while MS2b consisted of tidally locked binary stars with similar periods. There were four groups of giants. The G1/G3 stars were heavily spotted, tidally locked RS CVn stars with periods of tens
of days. The G2 stars were less luminous, heavily spotted, tidally locked, sub-subgiants
with periods $\sim 10$~days. The G4s group consisted of single stars with
luminosities intermediate to the G1/3 and G2 groups and slow rotation periods (approaching
$100$~days) that were almost certainly merger remnants.  The G1/G3 groups were combined because the G1 stars seemed to simply be G3 stars where the periodograms
favor finding $P_{rot}/2$. The G4b stars had similar
periods to the G4s stars, but consisted of sub-synchronously rotating binaries.

ASAS-SN uniformly studied the variability of all bright stars with a roughly daily cadence over a long
time baseline and a sensitivity of $\sim 0.03$~mag. Kepler (\citealt{keplerI2010}, \citealt{keplerII2010}),  by contrast,
provided exquisite photometric precision and high cadence observations of stars chosen to optimize 
planet searches in a small, fixed field off the Galactic plane. 
It was used to identify large samples of primarily
main sequence and sub-giant rotational variables (e.g., \citealt{McQuillan2014}, 
\citealt{Santos2019}, \citealt{Reinhold2023}). The \textit{Kepler} rotational variables
essentially all have amplitudes of $\ltorder 0.03$~mag that are disjoint from the
amplitudes of the ASAS-SN sample.
\cite{Daher2022} identified binaries in \textit{Kepler} using APOGEE estimates of the stellar rotation rate $v \sin i$ as a second variable. \cite{massarotti2008},
\cite{carlberg2011}, \cite{tayar2015} and \cite{Patton2023} also explored the binarity
of rapidly rotating spotted giants.  

The complementary properties of ASAS-SN and \textit{Kepler} highlight the potential value of comparing their
similarly sized populations of rotational variables. Here we carry out a similar analysis to 
\cite{phillips2023} for $\sim 50,000$ \textit{Kepler}
rotational variables from \cite{McQuillan2014} and \cite{Reinhold2023}. The latter includes samples considered by \cite{Santos2019}.
In Section \ref{sec:observations and methods}, we describe the data used in this 
study and its limitations. In Section \ref{sec:discussion}, we examine  
the distribution of the \textit{Kepler} rotational variables in class, binary properties (both fractions and 
tidal locking) and spot covering fractions in comparison to the   
\cite{phillips2023} sample.
Finally, in Section \ref{sec:conclusion}, we summarize and outline
future directions.

\section{Observations and Methods} \label{sec:observations and methods}

We started from 67,163 rotational variables from \textit{Kepler} \citep{keplerI2010,keplerII2010} with rotation periods from \cite{Reinhold2023}. \cite{Reinhold2023} did not include amplitudes, so we used amplitudes from the subset of these stars in \cite{McQuillan2014}. We cross-matched the stars to the \textit{Gaia} DR3 \citep{gaia2022} catalog, only keeping stars with \textit{Gaia} parallax signal-to-noise ratios $\varpi/\sigma_\varpi > 10$. We then matched these stars with the APOGEE DR17 \citep{apogeedr172022} and 2MASS \citep{2mass2006} catalogs. We used distance estimates from \cite{bailerjones2021} and the 3D {\tt `Combined 19'} {\tt MWDUST} \citep{bovy2016} model to estimate extinctions. These models are based on the \cite{drimmel2003}, \cite{marshall2006} and \cite{green2019} dust distributions. 

% We evaluated the binarity of the stars using the APOGEE DR17 and \textit{Gaia} DR3 catalogs, respectively. APOGEE provides higher precision radial velocity measurements but only 2,496 of the stars have multiple RV measurements (NVISITS $>$ 1), as summarized in Table \ref{table1}. \textit{Gaia} provides measurements for most of the \textit{Kepler} sample, but with much lower precision radial velocities. As in \cite{phillips2023}, we flagged APOGEE stars as binaries if the root-mean-square scatter of the velocities (VSCATTER) is $>$ $3$~km~s$^{-1}$, as they are almost certainly binaries \citep{badenes2018}. We flagged \textit{Gaia} stars as possible binaries using rv\underline{ }amplitude\underline{ }robust 
%  $> 20$~km~s$^{-1}$ and the criteria in \cite{katz2022}: 
% \begin{enumerate}
%     \item rv\underline{ }nb\underline{ }transits $\geq$ 5,
%     \item rv\underline{ }expected\underline{ }sig\underline{ }to\underline{ }noise $\geq$ 5 and, 
%     \item 3900 $\leq$ rv\underline{ }template\underline{ }teff $\leq$ 8000
% \end{enumerate}
% We also matched to the \textit{Gaia} DR3 binary catalogs (\citealt{Gosset2024}) to identify single- or double-lined spectroscopic binary (SB1, SB2) solutions so that we could compare the orbital and rotational periods.

We evaluated the binarity of the stars using the APOGEE DR17 and \textit{Gaia} DR3 catalogs. Identifying binary companions is important because short-period binaries can undergo tidal interactions that alter their rotation rates. For example, tidally interacting binaries cannot be used for gyrochronology. 
APOGEE offers high-precision radial velocities, but only 2,496 stars in our sample have multiple RV measurements (NVISITS $>$ 1), as summarized in Table \ref{table1}. Following \cite{phillips2023}, we flagged stars as binaries if their velocity scatter (VSCATTER) exceeded 3~km~s$^{-1}$, a conservative threshold indicating likely binarity \citep{badenes2018}. \textit{Gaia} provides radial velocity measurements for a larger fraction of the sample but at significantly lower precision. Nevertheless, we flagged possible binaries using the \texttt{rv\_amplitude\_robust} $> 20$~km~s$^{-1}$ criterion and the quality filters from \cite{katz2022}:
\begin{enumerate}
\item \texttt{rv\_nb\_transits} $\geq$ 5,
\item \texttt{rv\_expected\_sig\_to\_noise} $\geq$ 5, and
\item 3900K $\leq$ \texttt{rv\_template\_teff} $\leq$ 8000K
\end{enumerate}
These thresholds ensure that we detect binaries with significant dynamical effects while minimizing contamination from noise. We also cross-matched our sample with the \textit{Gaia} DR3 binary solutions from \cite{Gosset2024}, identifying both SB1 and SB2 systems. For these systems we can compare the rotational and orbital periods to see if systems are tidally locked.

Finally,
we used the star spot coverage estimates determined by the LEOPARD analysis of APOGEE spectra \citep{cao2022}. This algorithm fits the APOGEE spectrum using models with two different temperatures to estimate the temperature ratio
$X_{spot} = T_{in}/T_{out}$ between the regions in and out of spots
and the spot coverage fraction, $f_{spot}$. If the side
of the star we do not see has a spot fraction $f_{spot}'$, then the mean flux of the star satisfies
\begin{eqnarray}
  T_{eff}^4 &= & { 1\over 2} \left( f_{spot} T_{in}^4 + (1-f_{spot}) T_{out}^4 \right) \\
  &+  &{ 1\over 2} \left( f_{spot}' T_{in}^4 + (1-f_{spot}') T_{out}^4 \right) 
  \end{eqnarray}
  where $T_{eff}$ is the effective temperature.
The fractional change in flux relative to $T_{eff}^4$ between the two sides of
\begin{equation}
    \label{eq:1}
      v = { \Delta f \left(1-X_{spot}^4\right) \over 
        1 - \langle f \rangle (1-X_{spot}^4) }
\end{equation}
is a proxy for the variability due to the spots, where
$\Delta f = |f_{spot}-f_{spot}'|$ is the difference in
the spot fractions and 
$\langle f \rangle = (f_{spot}+f_{spot}')/2$
is the mean spot fraction. The actual variability also
depends on the stellar temperature and the observing
band passes, but this is an unnecessary complication since we simply want to illustrate the differences between the
samples.  We will use $v$ with 
$f_{spot}'=0$ ($\Delta f = f_{spot}$, $\langle f \rangle = f_{spot}/2$) to make the comparisons. One
important caveat for the LEOPARD estimates of spot properties is that they can be biased if the star
has a binary companion bright enough for the absorption lines of the companion to have an effect on the
estimates of $f_{spot}$ and $X_{spot}$.

\begin{figure*}
\centering
\includegraphics[width=.725\textwidth]{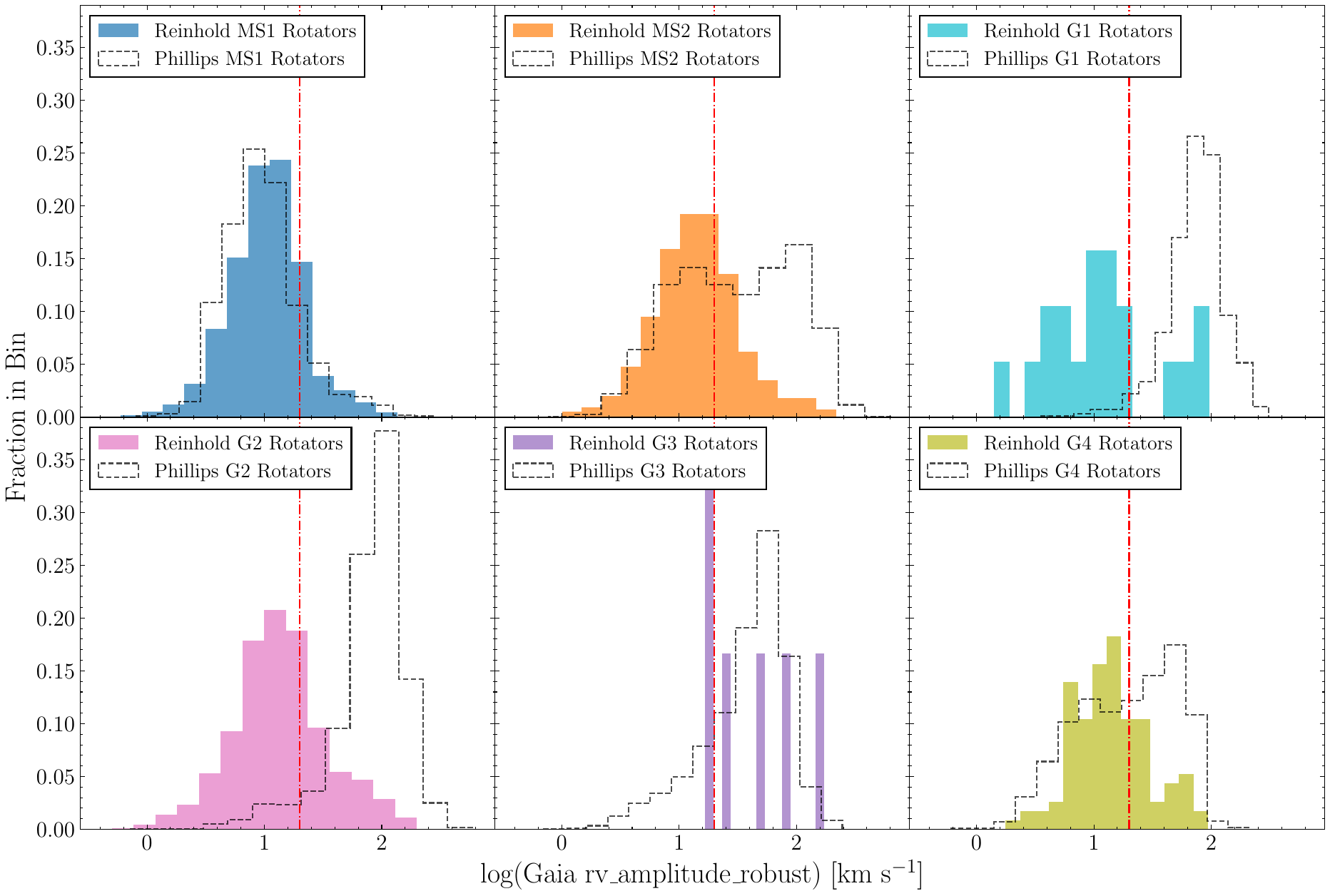}
\includegraphics[width=.725\textwidth]{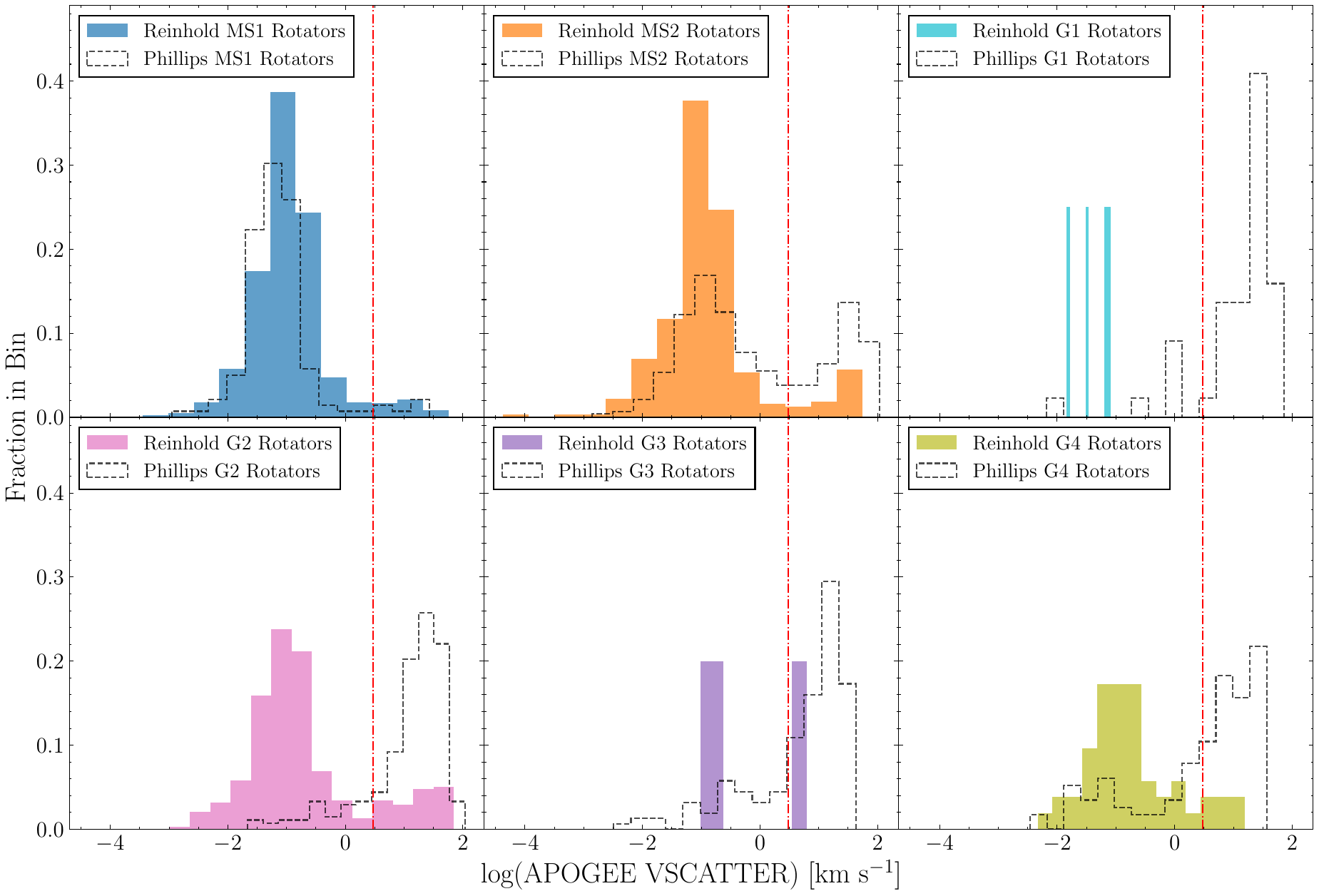}
\caption{Distributions of the MS1, MS2, G2, and G4 \textit{Kepler} stars in \textit{Gaia} rv\underline{ }amplitude\underline{ }robust (top panels) and APOGEE VSCATTER (bottom panels). Rotators to the right of the vertical lines at $20$~km~s$^{-1}$ (top panels) and $3$~km~s$^{-1}$ (bottom panels) are almost certainly binaries.
}
\label{fig:vhist}
\end{figure*}

\section{Discussion} \label{sec:discussion}

\begin{figure*}
    \centering
    \includegraphics[width=.8\textwidth]{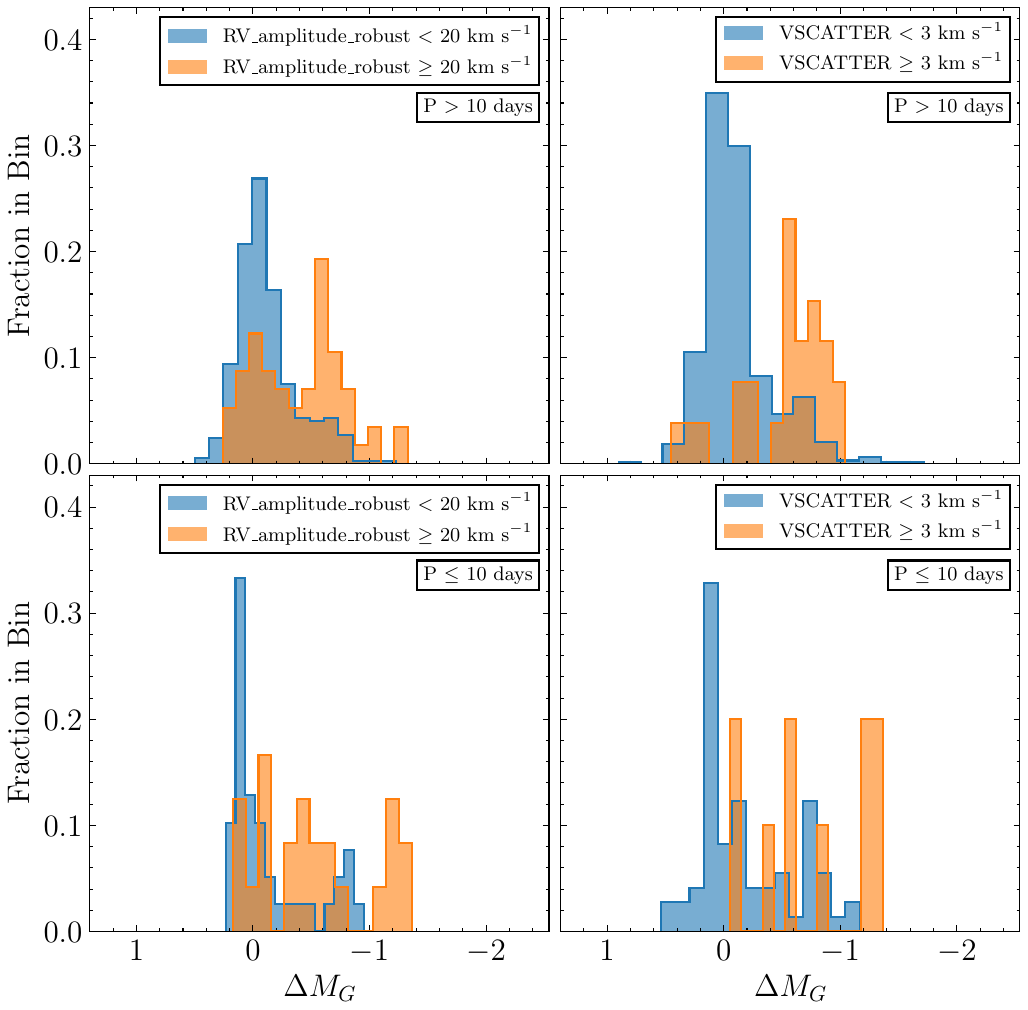}
    \caption{Distribution of the $\Delta M_G$ offset between a star's $M_G$ absolute magnitude and the three gigayear parsec isochrons \citep{parsec1, parsec2}. For both the MS1 and MS2 starts, the bottom (top) two panels display the distribution for stars with periods shorter (longer) than 10 days.}
    \label{fig:MgDist}
\end{figure*}

\begin{figure*}
    \centering
    \includegraphics[width=.8\textwidth]{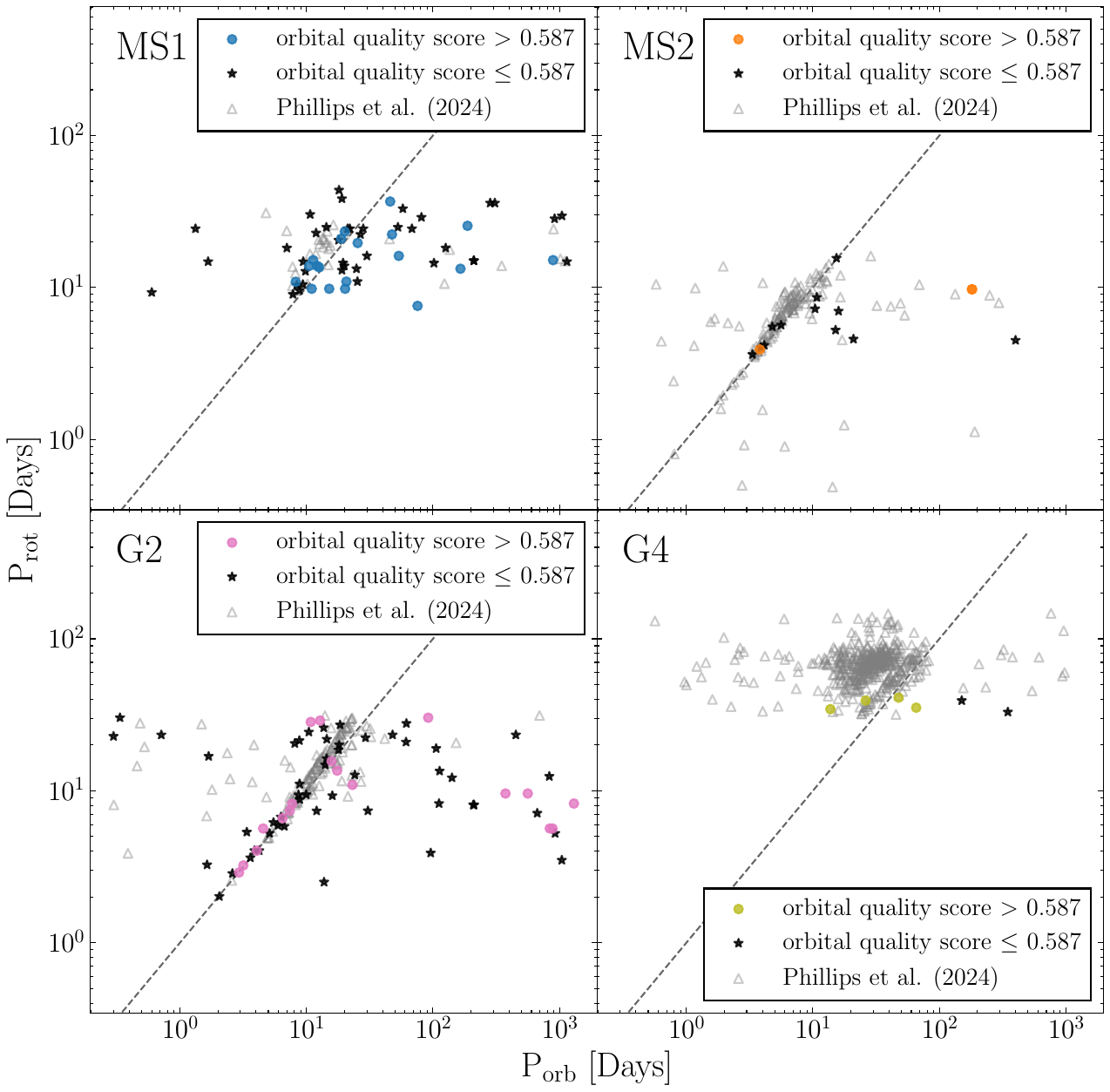}
    \caption{The \cite{Reinhold2023} rotational period $P_{rot}$ compared to the \textit{Gaia} orbital period $P_{orb}$.
    Tidally locked systems would lie on the dashed line with $P_{rot} = P_{orb}$. The orbital quality scores use the criteria in \cite{Bashi2022}}
    \label{fig:locked}
\end{figure*}

Figure \ref{fig:cmd} shows the distributions of a random 15 percent of the stars in $M_{K_s}$ and period and in a \textit{Gaia} color-magnitude diagram (CMD). The stars are divided into the same classes as \cite{phillips2023} using the boundaries shown in Figure \ref{fig:cmd}. Table \ref{table1} gives the number of rotational variables from \cite{Reinhold2023} with \textit{Gaia} magnitudes and parallax errors $\sigma_\varpi/\varpi<1/10$
divided into these groups. The \textit{Kepler} samples
focused on main sequence stars and, to a lesser extent, sub-giants, but generally avoided stars on the giant
branch, as seen in Figure \ref{fig:cmd}. The \textit{Kepler} samples also discriminated against wider binaries, avoiding resolved systems or systems
with large \textit{Gaia} RUWE (Renormalized Unit Weight Error) where the astrometry residuals suggest the presence of a companion.  There does not
seem to be any significant discrimination against short period binaries (e.g. avoiding systems lying above the
main sequence). \cite{Reinhold2023} only allowed rotational periods shorter than 50 days leading to the sharp edge in the period distribution. For these reasons, we focus on the 
MS and sub-giant systems with limited results
for the more luminous giants.

The distribution of the MS stars in the left panel of Figure ~\ref{fig:cmd} is strikingly different from \cite{phillips2023}. Both samples have a clump of MS1 stars with a fairly well defined peak near a period of 25~days.  But in the \textit{Kepler} sample, the MS2 group is essentially absent and lacks a well defined peak. 

In \cite{phillips2023}, only 14\% of the MS stars belonged
to the MS1 class, while here 88\% of them belong to MS1.  This cannot simply be an unrecognized bias against short period binaries in Kepler.  The MS1 and MS2 binary fractions in \cite{phillips2023} were 4.3\% and 34.3\%, respectively,
and if we correct our current numbers for losing these fractions, the corrected ratio would be that 83\% of the Kepler
stars are in MS1.   Moreover, the MS2 systems have shorter periods ($<$10 days) than the MS1 systems and the MS2 binaries
in \cite{phillips2023} were tidally locked, so any biases against binaries in \textit{Kepler} would do little to their inclusion.

Figure \ref{fig:vhist} shows the distribution of the MS1, MS2, G2 and G4 classes in \textit{Gaia} rv\underline{ }amplitude\underline{ }robust and APOGEE VSCATTER as compared to the equivalent distributions from \cite{phillips2023}. The MS1 rotators are again overwhelmingly dominated by single stars with VSCATTER and rv\underline{ }amplitude\underline{ }robust distributions that are very similar to the MS1 rotators in \cite{phillips2023}. The \textit{Kepler} MS2 distributions are very different, with much lower binary fractions than in \cite{phillips2023}. While the \textit{Kepler} MS2 systems have a tail to higher velocity scatters, only 8.9\% of the systems have a
$\hbox{VSCATTER}>3$~km/s as compared to 34.3\% in the \cite{phillips2023} sample.  That the distribution shows
a peak at the highest values of VSCATTER strongly suggests
that there are real binaries in the \textit{Kepler} MS2 group. If we define $x \equiv v_{obs}/v_{max}$ as the ratio of the scatter that would be observed edge on to
the scatter $v_{obs}=v_{max}\sin i $ that would be observed at inclination $i$, then the probability distribution of $x$ for
random inclinations is $x(1-x^2)^{-1/2}$ which peaks for $v_{obs} \rightarrow v_{max}$, similar to what we see in
Figure ~\ref{fig:vhist}.

\cite{simonian2019} examined the binary fraction of the rapidly rotating ($1.5<P<7$~day) \cite{McQuillan2014} rotational variables
using magnitude offsets from the main sequence.  Figure ~\ref{fig:MgDist} shows a version of this using the offset of fixed color ($\Delta M_g$) between
the MS1/MS2 stars and a 3~Gyr PARSEC \citep{parsec1, parsec2} isochrone for the stars which can be flagged as binaries using either \textit{Gaia} or APOGEE velocity
scatters.  If the distribution of mass ratios $q$ is uniform, and the main sequence luminosity scales as mass $M^a$ with $a=3$-$4$,
the distribution in $x=\Delta M_g$ is
\begin{equation}
      {d P \over dx} \propto 10^{-0.4x} \left( 10^{-0.4x}-1 \right)^{1/a-1}
\end{equation}
is quite flat despite the singularity if smoothed over even $\Delta M_g = 0.1$.  For $a = 3$, the binary fraction rises by roughly a factor of two from nearly equal-mass binaries ($\Delta M_g = -1.5$) to systems with low-mass companions ($\Delta M_g \rightarrow 0$). Given the crudeness of our estimate of $\Delta M_g$ (see \citet{simonian2019} for a more detailed treatment) and the modest sample size, this is broadly consistent with the binary distributions shown in Figure ~\ref{fig:MgDist}. While \citet{simonian2019} report a significantly higher binary fraction (59\%) than we find for the main sequence population as a whole, their analysis focused on very rapidly rotating stars.  Our binary fractions also are not corrected for completeness due to inclination effects and the RV sampling simply missing phases with large velocity differences.
The first effect is small even if the true VSCATTER is only modestly above the $3$~km~s$^{-1}$ limit (it includes $>90\%$  of systems with
equatorial $VSCATTER = 6$~km/s).  The effects of the phase sampling are likely larger for smaller numbers of APOGEE visits, but a full statistical correction of the phase sampling would be a significant Monte Carlo modeling project.
The panels in Figure ~\ref{fig:MgDist} separate the distributions by rotation period, highlighting this distinction: among stars with $P \leq 10$ days, we find a binary fraction of 12.1\% (38.1\%) using VSCATTER (rv\underline{ }amplitude\underline{ }robust), compared to just 3.9\% (13.3\%) for the slower rotators. This trend is qualitatively consistent with the findings of \citet{simonian2019}, who also observed a higher binary fraction among the most rapidly rotating systems.

\begin{figure*}
\centering
\includegraphics[width=.8\textwidth]{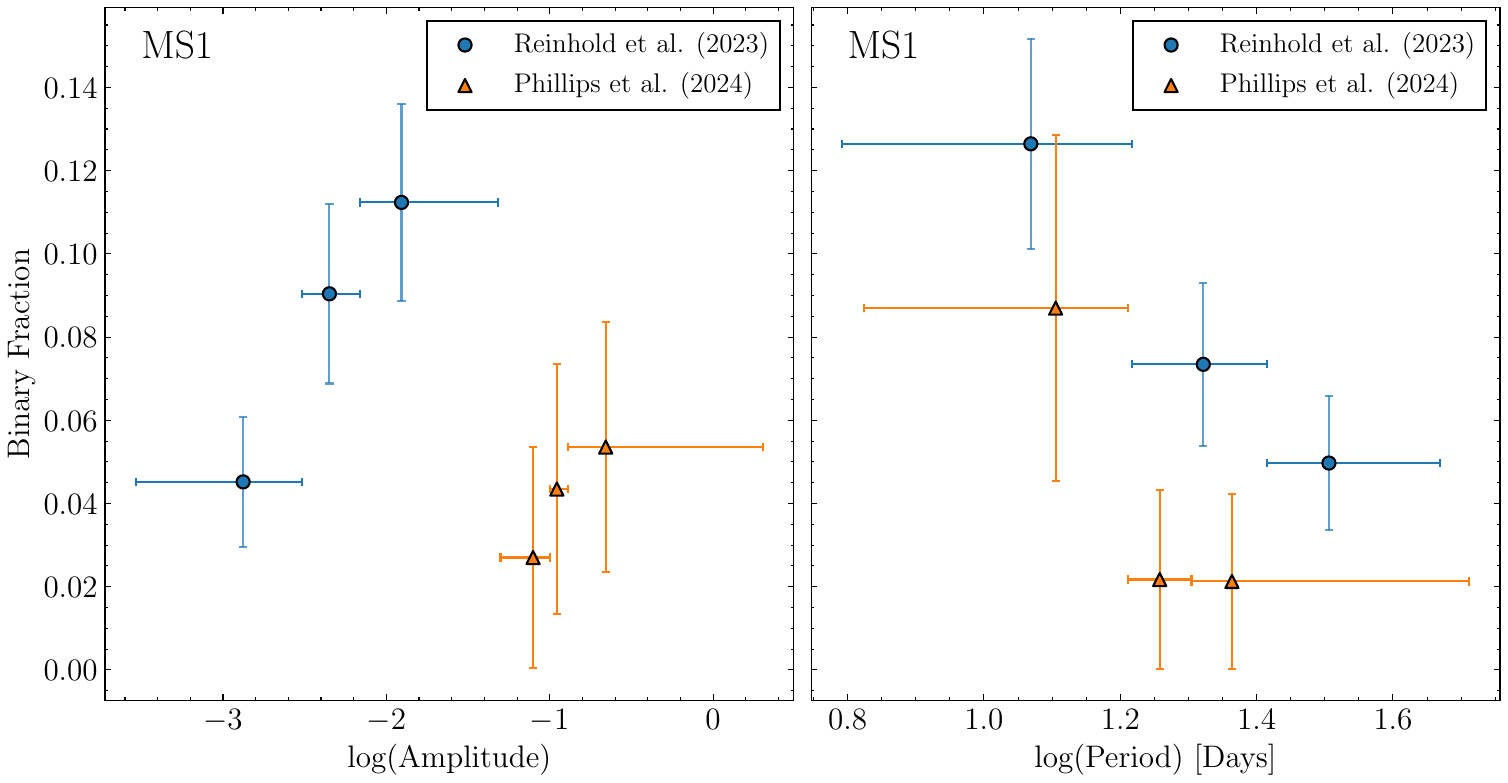}
\includegraphics[width=.8\textwidth]{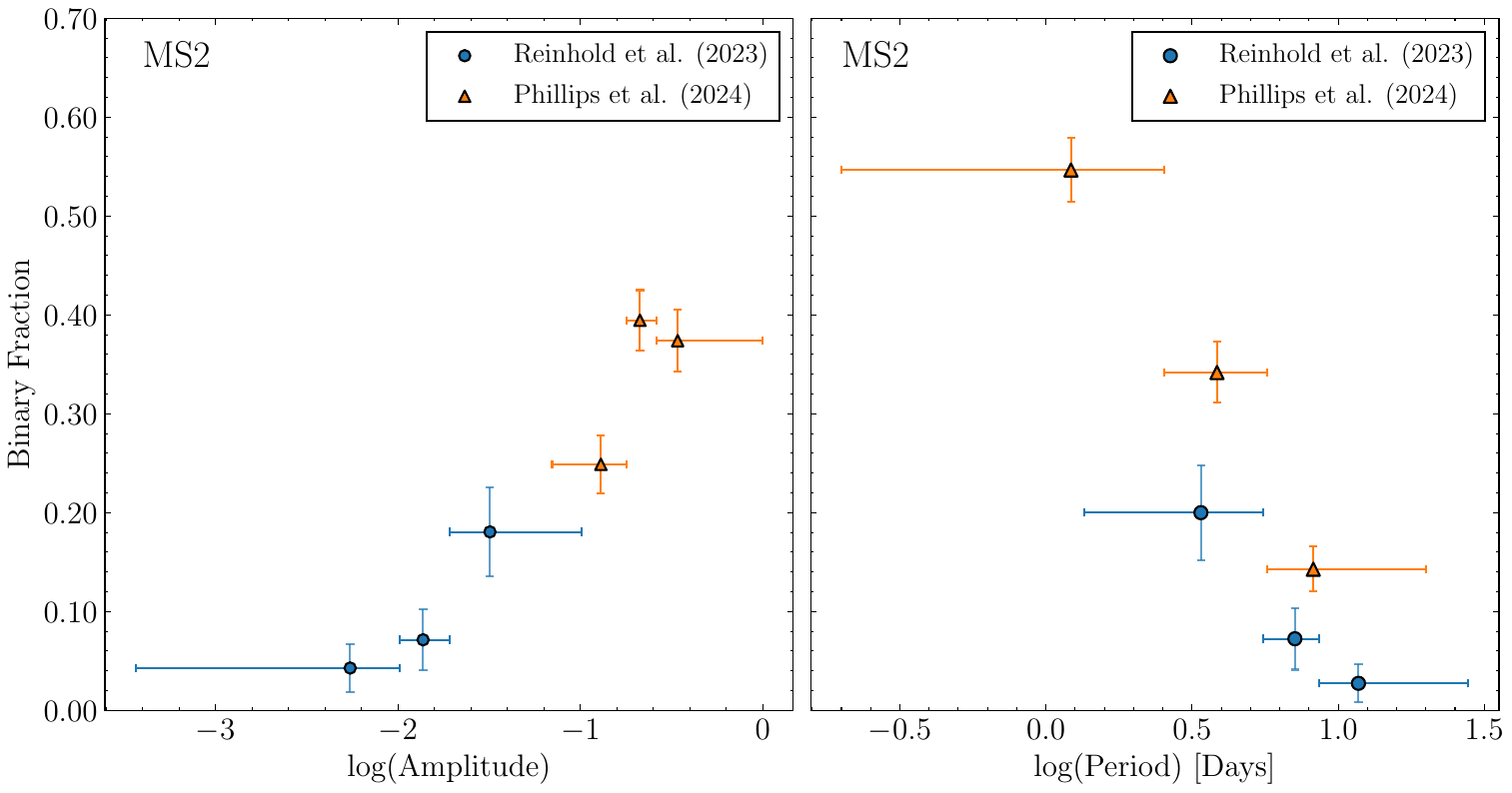}
\caption{Binary fractions of MS1 (top panels) and MS2 (bottom panels) rotational variables from both \textit{Kepler} and \cite{phillips2023} in amplitude (left) and period (right). Each amplitude and period bin contains an equal number of stars.}
\label{fig:amplitude}
\end{figure*}

\begin{figure*}
    \centering
    \includegraphics[width=.8\textwidth]{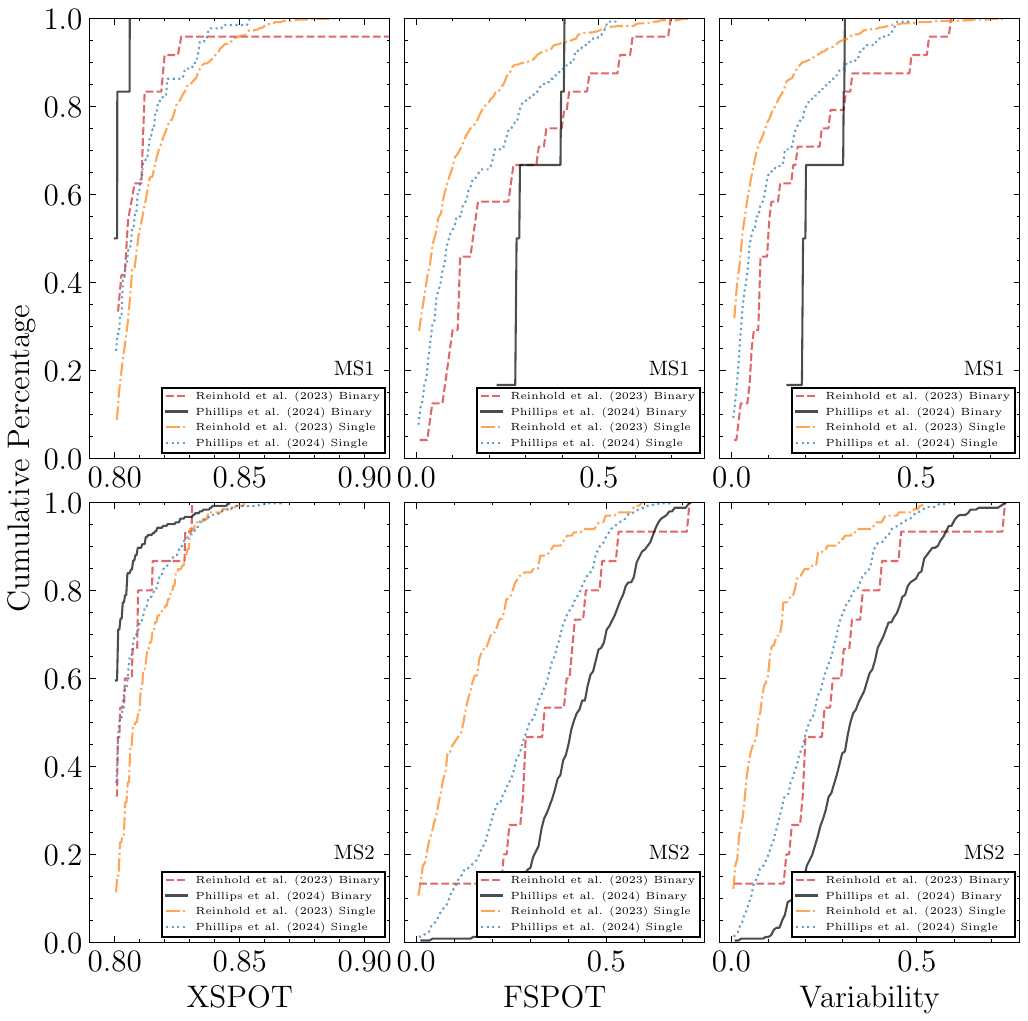}
    \caption{Integral distributions of the XSPOT, FSPOT, and variability $v$ (Eqn. \ref{eq:1}) for the main sequence \cite{Reinhold2023} and \cite{phillips2023} binaries and single stars. This variability roughly corresponds to the fractional amplitude if only the ``observed" side of the star is spotted.}
    \label{fig:variability}
\end{figure*}

In \cite{phillips2023}, the lower luminosity G2 and G4 classes are dominated by binaries, although G4 had a minority
population of single stars. As we see in Figure \ref{fig:vhist}, this is completely reversed for the \textit{Kepler} stars. Both groups are dominated by
single stars with a minority of binaries.  For example 84.9\% (63.5\%) of G2 (G4) stars had $\hbox{VSCATTER}>3$~km/s
in \cite{phillips2023}, while only 16.1\% (11.5\%) of the \textit{Kepler} stars have such large VSCATTER values.  For the more
luminous giants, G1 has a sufficient number of \textit{Gaia} RV\underline{ }amplitude\underline{ }robust measurements to make a clear comparison, and we find that the \textit{Kepler} stars are primarily single stars while the \cite{phillips2023} stars are mostly binaries.
For the G3 giants, there are too few \textit{Kepler} stars with velocity information to make a comparison.

In addition to the abundance of binaries, we are also interested in whether the systems are tidally locked, with
rotational periods $P_{rot}$ equal to their orbital periods $P_{orb}$.  We can make the comparison for the limited
number of systems with orbital solutions in \textit{Gaia} DR3 (\citealt{Gosset2024}) with the caveat that we have found significant fractions
of period mismatches between ASAS-SN photometric periods and \textit{Gaia} DR3 orbital periods in eclipsing binaries (e.g, \citealt{Rowan2023}).
Here we use the criteria of \cite{Bashi2022} to flag potentially bad orbital solutions.  Figure ~\ref{fig:locked} 
compares the two periods for the MS1, MS2, G2 and G4 groups - there are too few G1 and G3 stars with orbital
solutions to make a comparison.  

 \cite{phillips2023} had too few MS1 systems with \textit{Gaia} orbits to draw strong conclusions, while the binary MS2 systems were dominated by tidally synchronized systems. For the \textit{Kepler} stars, the majority of MS1 binary  systems are not tidally synchronized and the orbital periods are much longer than the rotational periods. The gap in the orbital periods near one year is due to the \textit{Gaia} scanning
 patterns. There does seem to be a minority of systems clustered near, but not on, the $P_{orb}=P_{rot}$ line.  This scatter
 is likely real since in the \cite{phillips2023} samples (and for G2 here) there are populations where the two periods are
 much more similar.  
 The MS2 binaries are also dominated by non-synchronous systems, but now with a minority lying on or close to the $P_{orb}=P_{rot}$ line. In both samples, the periods of the synchronized systems are $\sim 10$ days or lower.
The G2 systems in \cite{phillips2023} were essentially all tidally synchronized binaries, while the G4 systems were dominated by sub-synchronous ($P_{rot} > P_{orb}$) systems. Only $\sim1/4$ of the \textit{Kepler} G2 systems are (close to) synchronous, while none of the G4 systems are synchronous. The G4 systems also almost all have $P_{orb} > P_{rot}$ rather than the reverse.

As discussed earlier, there is an enormous difference in the variability amplitudes of the ASAS-SN and \textit{Kepler} samples. Figure \ref{fig:amplitude} shows the APOGEE binary fractions of the \textit{Kepler} and ASAS-SN MS1 and MS2 stars in amplitude and period.  As noted above, the binary fractions are not corrected for completeness, but the corrections should be
roughly the same in each period bin. For both classes and populations, binary fractions increase with amplitude and decrease with period. The MS1 rotators from \textit{Kepler} generally have higher binary fractions  than the MS1 rotators from \cite{phillips2023} despite their lower amplitudes, and this is also seen for each of the overlapping period bins. The MS2 rotators show a steadily increasing binary fraction with amplitude across both samples. At fixed periods, the binary factions of the higher amplitude rotators are larger, and the \textit{Kepler} population has none of the very short period (sub-day) systems seen in the \cite{phillips2023} sample. There were not enough systems in the giant classes
to do a similar analysis.

Figure \ref{fig:variability} presents the integral distributions of the MS1 and MS2 groups in spot coverage fraction ($f_{spot}$), temperature ratio ($X_{spot}$) and variability ($v$, Eqn.~\ref{eq:1}) for the \textit{Kepler} and \cite{phillips2023} samples.  The ASAS-SN systems generally have larger
temperature differences and larger spot fractions, leading to larger estimates of the variability.  For the MS1 group, the temperature ratios are
quite similar between the samples, with larger differences in the spot fractions.  For the MS2 group, the differences in both temperature and spot fractions are larger. The binary systems, are more spotted in both samples, with the caveat that the \cite{cao2022} estimates can be skewed by the spectral contributions of the secondary.

% While the ASAS-SN variables are predicted to be more variable based on the spot models, the differences in the estimated
% amplitudes are significantly smaller than the observed order of magnitude amplitude difference.  The amplitude estimate used for 
% Figure ~\ref{fig:variability} took the limiting case where only the observed side of the star is spotted (i.e., $f_{spot}'=0$).  The
% simplest way to greatly increase the amplitude difference is for the \textit{Kepler} rotational variables to not only have smaller temperature
% differences and spot coverage fractions, but to also have spots more uniformly distributed over the surface of the star.  In the
% simple model associated with Eqn.~\ref{eq:1}, the \textit{Kepler} stars also need to have $f_{spot} \simeq f_{spot}'$ so that $\Delta f$ is small,
% while the ASAS-SN stars have significantly non-uniform spot distributions.  If we view this simply as a problem of randomly distributing
% $N$ spots on two ``sides'' of a star, the median fractional difference in the number of spots is 
% $\Delta f /\langle f \rangle=0.4$, $0.13$, $0.08$, $0.04$ and $0.02$ for $N=10$, $30$, $100$,  $300$, and
% $1000$ spots, so the difference in amplitudes requires not only the observed differences in temperature and spot covering fraction,
% but also that the spotted \textit{Kepler} stars have an order of magnitude more spots than the ASAS-SN stars.

While the ASAS-SN stars are predicted to be more variable based on the spot models, the modeled amplitude differences remain significantly smaller than the observed order-of-magnitude difference in variability. The estimates used in Figure~\ref{fig:variability} assume a limiting case where only the observed side of the star contains spots ($f_{spot}' = 0$). To reconcile this with the much lower observed amplitudes in the \textit{Kepler} sample, the stars would need not only smaller spot coverage fractions and lower temperature contrasts, but also more uniformly distributed spots across their surfaces. In the framework of Eqn.~\ref{eq:1}, this corresponds to $f_{spot} \simeq f_{spot}'$, yielding a small $\Delta f$ and thus reduced photometric variability.

To illustrate this, we considered a simple toy model in which $N$ identical spots are randomly distributed across two hemispheres of a star. The fractional difference in spot coverage between the hemispheres, $\Delta f/\langle f \rangle$, decreases with increasing $N$, approximately scaling as $1/\sqrt{N}$ for large $N$. For example, the median fractional asymmetry is $\Delta f/\langle f \rangle \sim 0.4$ for $N = 10$ spots, but drops to $\sim 0.04$ for $N = 300$ spots. Thus, to maintain the low amplitudes observed in the \textit{Kepler} sample, stars would either require an order of magnitude more (and likely smaller) spots than typical ASAS-SN stars, or the true spot distributions must be more uniform than assumed.
There are limits to this argument since there is evidence that spot distributions in close binaries 
may not be random and can exhibit latitudinal or longitudinal preferences \citep{sethi2024tightstellarbinariesfavour}. 

\section{Conclusion}
\label{sec:conclusion}

As discussed in the introduction, \cite{phillips2023} analyzed the clusters of rotational variables in absolute magnitude
and period found by \cite{christy2023} for their binary properties and star spot coverage fractions.  Here we carry out
a similar analysis of the much lower variability amplitude rotational variables from Kepler, using the results from 
\cite{Reinhold2023} and \cite{McQuillan2014}.  The \textit{Kepler} samples contain very few of the higher luminosity giant
rotator groups G1 and G3, so we focus on the MS groups and the less luminous G2 and G4 giant groups.  In general,
there are significant differences between the two samples for all of the groups.

\textbf{MS1}: The first striking result is that the longer period MS1s group dominates the \textit{Kepler} sample, with 88\% of the main sequence stars, compared to only 14\% in \cite{phillips2023}. These stars are predominantly single in both samples and have modest binary fractions ($\ltorder 10\%$) that increase with variability amplitude and decrease with rotation period. The Kepler
sample has modestly higher binary fractions despite the lower amplitudes, but typically at longer periods. The majority of the \textit{Kepler} binaries in this group are far from tidal synchronization.

% The \textit{Kepler} stars have lower spot coverage fractions and slightly smaller temperature contrasts. These differences are not large enough to explain the large amplitude differences. This implies that the \textit{Kepler} stars must have many more spots in order to further suppress their variability.

The \textit{Kepler} stars exhibit lower spot coverage fractions and slightly smaller temperature contrasts compared to the ASAS-SN sample. However, these differences are not sufficient to explain the significantly lower variability amplitudes observed in \textit{Kepler} light curves. This implies that many of these stars must host a greater number of more symmetrically distributed spots, which would suppress variability without requiring low total spot coverage.

Several factors likely contribute to this discrepancy. The \textit{Kepler} field was chosen to contain an older stellar population, making rapidly rotating, highly active stars intrinsically rare in the sample. Stellar rotation also slows substantially over time, so the majority of stars in the \textit{Kepler} field are expected to be older and magnetically inactive, with correspondingly low spot coverage. In contrast, ASAS-SN is biased toward detecting high-amplitude variability in bright stars, favoring the inclusion of more active systems with asymmetric spot distributions. Moreover, the \textit{Kepler} target selection prioritized stars suitable for detecting transiting exoplanets, and this process may have implicitly excluded or de-emphasized active stars, further shaping the sample's variability properties.

\textbf{MS2}: In the \cite{phillips2023} sample, the MS2 group has comparable numbers of MS2s single stars and MS2b binary
stars, with the binaries largely being tidally synchronized.  For the \textit{Kepler} stars, there are many fewer binaries and most
are not tidally synchronized.  The binary fraction is less than 1/2 that of the \cite{phillips2023} sample, with a smooth
trend of binary fraction with variability amplitude across both samples (Figure \ref{fig:amplitude}). The temperature contrasts and spot fraction differences (Figure \ref{fig:variability}) are larger for the MS2 stars, but the amplitude differences again seem to require that the \textit{Kepler} stars have many more spots.

\textbf{G2 and G4}: In \cite{phillips2023}, these sub-giant stars are essentially all tidally locked binaries. In \textit{Kepler} they are mostly single stars, while the binaries are a mixture of tidally locked binaries and systems with $P_{orb} \gg P_{rot}$. These sub-giant groups show pronounced differences between the two samples. The \cite{phillips2023} G4 systems are mostly
sub-synchronously rotating binaries with a modest fraction of single stars. In \textit{Kepler} they are almost all single stars,
and those in binaries have orbital periods uncorrelated with the rotation periods.  There were too few \textit{Kepler} stars with
spot fraction estimates to make a comparison.

Since tidally locked stars generally have short rotational periods, it is not surprising that the high variability
amplitude ASAS-SN systems systematically show larger binary populations than the low amplitude \textit{Kepler} systems.  
However, APOGEE is sensitive to quite wide binaries ($\sim 3000$~days, see \citealt{Daher2022}), and most of
this period range would not lead to tidal interactions. The spot statistics for the binaries should be interpreted with caution, however, because spectral lines from the convergence can skew the spot statistics. The \textit{Kepler} stars are generally non-synchronous binaries, particularly in the MS1 and MS2 classes, with synchronization restricted to shorter orbital periods ($\lesssim10$ days). This contrasts with the tidal synchronization observed in many \cite{phillips2023} binaries.

Across both samples, higher variability amplitudes correspond to increased binary fractions, underscoring the interplay between binarity, magnetic activity, and rotational evolution. Binary systems have higher spot coverage fractions ($f_{spot}$) but, the \textit{Kepler} binaries have slightly lower values than their counterparts in \cite{phillips2023}.

There is considerable scope for expanding population studies of rotational variables.  First, there will be larger samples
of low amplitude rotational stars identified by TESS (Transiting Exoplanets Survey Satellite) (e.g., \citealt{Claytor2022}, \citealt{Holcomb2022}, \citealt{Colman2024}),
with the caveat that the TESS observing strategy makes it difficult to identify longer period ($\gtorder 2$~weeks) systems.
Ground based variability surveys like ASAS-SN will continue to expand the samples of higher amplitude systems, and will be
able to identify lower amplitude systems as the light curves extend in time. 

The real limitation on population
studies is the small fraction with good binary discrimination or binary orbit solutions.  While the \textit{Gaia} rv\underline{ }amplitude\underline{ }robust is available for most of these stars, it is only sensitive to scatters $>20$~km/s, and
only a tiny fraction of the systems have full \textit{Gaia} orbital solutions.  Fortunately, the number of systems with orbital
solutions should expand by over an order of magnitude with \textit{Gaia} DR4.   SDSS Milky Way Mapper (see \citealt{Kollmeier2017})
will greatly increase the number of systems with APOGEE's higher precision velocity scatters.  As shown by 
\cite{phillips2025}, the velocity scatter distributions can be used to statistically estimate
the primary masses and secondary mass ratios even without having a full orbit solution.

\begin{acknowledgments}
CSK is supported by NSF grants AST-2307385 and AST-2407206. AP is supported by the National Science Foundation Graduate Research Fellowship under Grant No. DGE 2140743. MHP and LC acknowledge support from NASA grant 80NSSC23K0205.
\end{acknowledgments}


\vspace{5mm}




\clearpage
\bibliography{ref}{}
\bibliographystyle{aasjournal}



\end{document}